\documentclass[fleqn,usenatbib]{mnras}

\usepackage{newtxtext,newtxmath}

\usepackage[T1]{fontenc}

\DeclareRobustCommand{\VAN}[3]{#2}
\let\VANthebibliography\thebibliography
\def\thebibliography{\DeclareRobustCommand{\VAN}[3]{##3}\VANthebibliography}

\usepackage{graphicx}	
\usepackage{amsmath}	

\newcommand{\cow}{{AT\,2018cow}}

\title[AT~2018cow at $\sim$ 5 years]{AT~2018cow at $\sim$5 years: additional evidence for a tidal disruption origin}

\author[A. Inkenhaag et al.]{
Anne Inkenhaag,$^{1}$\thanks{E-mail: ai707@bath.ac.uk}
Andrew J. Levan,$^{2,3}$
Andrew Mummery,$^{4}$
Peter G. Jonker$^{2}$
\\
$^{1}$Department of Physics, University of Bath, Claverton Down, Bath BA2 7AY, UK\\
$^{2}$Department of Astrophysics/IMAPP, Radboud University, 6525 AJ Nĳmegen, The Netherlands\\
$^{3}$Department of Physics, University of Warwick, Gibbet Hill Road, Coventry, CV4 7AL, UK \\
$^{4}$Oxford Theoretical Physics, Beecroft Building,  Clarendon Laboratory, Parks Road, Oxford, OX1 3PU, United Kingdom
}

\date{Accepted 2025 October 07. Received 2025 September 17; in original form 2025 June 02}

\pubyear{\the\year{2025}}

\begin{document}
\label{firstpage}
\pagerange{\pageref{firstpage}--\pageref{lastpage}}
\maketitle

\begin{abstract}
The Luminous Fast Blue Optical Transient (LFBOT) AT~2018cow is the prototype of its class with an extensive set of multi-wavelength observations. Despite a rich data set there is, still, no consensus about the physical nature and origin of this event. AT~2018cow remained UV bright 2--4 years after the explosion, which points at an additional energy injection source, most likely from an accretion disk. We present additional late time UV data obtained with the \textit{Hubble Space Telescope}, to show there is no significant fading in the optical since the last epoch and only marginal fading in the UV. The new UV data points match the predictions of previously published accretion disk models, where the disk is assumed to form from the tidal disruption of a low mass star by an intermediate mass black hole. This consistency provides evidence that AT~2018cow could indeed be a tidal disruption event. The marginal decay is in contrast with the predictions of light curves produced by interacting supernovae. The difference between expectations for disc emission and interacting supernovae will further increase with time, making data at even later times a route to robustly rule out interaction between supernova ejecta and circumstellar material.
\end{abstract}




\section{Introduction}

Luminous fast blue optical transients (LFBOTs) are a relatively recently discovered class of multi wavelength transients \citep{Prentice2018, Perley2019, Coppejans2020, Ho2020, Perley2021, Yao2022, Matthews2023, Chrimes2024, Ho2024, Pursiainen2025}. These events are characterised by fast rise and decay times, high peak luminosities (absolute peak magnitude $M \lesssim -19$), and early spectra dominated by a blue featureless continuum. There are currently less than ten LFBOTs known. 

The prototypical LFBOT is \cow\ \citep{Prentice2018, Perley2019} which was detected and reported rapidly, and is the closest known LFBOT at a distance of just 63.0$\pm$4.4~Mpc \citep[$z=0.01404\pm0.00002$, SDSS DR6;][]{SDSS2008}. This combination triggered an extensive, multi-wavelength follow-up campaign, resulting in \cow\ being one of the most well-studied transients to date. Other LBFOTs were either at greater (sometimes cosmological, $z>0.1$) distances (e.g., AT2020mrf; \citealt{Yao2022} and AT2023fhn; \citealt{Chrimes2024}) or were not immediately recognised as LFBOTs (e.g., CSS161010 which was reported in 2016, but only published in 2020; \citealt{Coppejans2020, Gutierrez2024}), meaning there was less opportunity to obtain such extensive data as exists for \cow.

The ultimate physical origin of LFBOTs remains uncertain, despite much progress being made in the last few years. Among a wide range of theoretical models seeking to explain the origin of these events, the community has converged onto a much-reduced subset of models: the `naked' engine model, where the emission comes from accretion onto a compact object formed during a core-collapse supernova with small ejecta mass \citep[e.g.,][]{Prentice2018, Perley2019, Margutti2019, Mohan2020}, an intermediate mass black hole (IMBH) tidally disrupting a white dwarf (WD) \citep[e.g.,][]{Kuin2019,Perley2019}, or a low-mass main sequence star \citep{Inkenhaag2023}, the tidal disruption of a Wolf-Rayet star by a stellar mass black hole \citep{Metzger22b}, or interaction between an outflow and a dense circumstellar medium \citep[e.g.,][]{FS2019, Leung2020, Xiang2021, Pellegrino2022a}. Both a compact object \citep[e.g.,][]{Margutti2019, Pasham2021} and CSM \citep[e.g.,][]{Coppejans2020} have been observationally shown to be present in these systems, providing observational justification to consider the models as well. Furthermore, \citep{Gutierrez2024} provide spectroscopic evidence in support of the TDE scenario for CSS161010. The model convergence to the aforementioned sub-set of models is due to both the discovery of more LFBOTs in recent years and early time observations across the electromagnetic spectrum, but specifically due to \textit{HST} UV detections of the archetypal LFBOT AT~2018cow at late times \citep{Sun2022, Sun2023, Chen2023, Inkenhaag2023}, which ruled out many promising model candidates.

These late-time \textit{HST} detections demonstrated that the UV emission of AT~2018cow remained several magnitudes brighter at $\Delta t \sim 700-1450$ days than the extrapolated early time emission \citep{Sun2022}, and varied only slowly in this phase \citep{Sun2023}. While \cite{Sun2022} argued this late time emission was likely not from accretion onto a compact object, arguing that the plateau phase in the light curve would require delayed accretion on the object, \cite{Inkenhaag2023} instead showed that the accretion disk formed at early times by a M$_{\rm BH} = 10^{3.2\pm0.8}$~M$_\odot$ black hole tidally disrupting a low mass star could, in fact, readily explain the observed late time properties, rejuvenating the tidal disruption event (TDE) model. 

Indeed, bright long-lived  UV emission is a ubiquitous feature of tidal disruption events \citep{VanVelzen2019, MumBalb20a, Mummery_et_al_2024}. Every classical (galactic centre supermassive black hole) TDE which is nearby ($z<0.07$) and has more than 3 years of optical/UV data has been detected to undergo this late time ``plateau phase'' in the UV (for a total of $N=40$ sources) \cite{Mummery_et_al_2024, MummeryVV24}. The amplitude of this plateau is known (theoretically and observationally) to correlate with the central black hole mass with a $L_{\rm plat} \propto M_{\rm BH}^{2/3}$ scaling \citep{Mummery_et_al_2024}. It is this scaling which, when applied to the low luminosity of AT~2018cow at late times, implies an intermediate mass black hole \citep{Inkenhaag2023}. 

In this work we present an additional set of multi-wavelength observations of \cow\ at $\sim1900$~days, using \textit{HST}/WFC3 in four filters. In Section~\ref{sec:data} we describe the data and the analysis. We also present our results in this section. Section~\ref{sec:discussion} contains the discussion of the results in the context of the progenitor models of \cow, and we summarise our findings and conclusion in Section~\ref{sec:conclusions}. 

All magnitudes are presented in the AB magnitude system unless specified otherwise, with 1$\sigma$ errors. Upper limits are defined as 3$\sigma$ upper limit. Throughout the paper we use H$_0 = 67.8$\,km\,s$^{-1}$\,Mpc$^{-1}$, $\Omega_{\rm m}=0.308$ and $\Omega_{\rm \Lambda} = 0.692$ \citep{Planck2016}.

\section{Data and analysis} \label{sec:data}

\begin{table*}
\caption{Result of aperture and PSF photometry on the latest epoch of observations of \cow\ at 1900 and 2043~days. Aperture photometry was done using a circular aperture of r=0.08~arcsec, while PSF photometry was done using {\sc dolphot}. The magnitudes are corrected for Galactic extinction (A$_{\rm F225W} = 0.524$, A$_{\rm F336W} = 0.334$, A$_{\rm F555W} = 0.214$, A$_{\rm F814W} = 0.115$) and include aperture correction. 
}
\label{tab:data}
\hspace*{-1.0cm}\begin{tabular}{ccccccccccccc}
\hline
 Filter & Epoch  & \# of &Total Exp.~time & Aperture phot.&  Aperture phot. & {\sc dolphot} & {\sc dolphot}\\
 & (day) & exp. & (sec) & F$_\nu$ ($\mu$Jy) &  (mag) & F$_\nu$ ($\mu$Jy) & (mag) \\
\hline
F225W & 1900 & 3 & 2019 & $0.64\pm0.03$ & $24.38\pm0.06$ & $0.60\pm0.04$ & $24.45\pm0.06$ \\
F336W & 1900 & 3 & 1944 & $0.49\pm0.02$ & $24.68\pm0.05$ & $0.38\pm0.02$ & $24.96\pm0.05$\\
F555W & 1900 & 3 & 1437 & $0.27\pm0.01$ & $25.31\pm0.05$ & $0.17\pm0.01$ & $25.83\pm0.07$\\
F814W & 2043 & 2 & 1214 & $0.22\pm0.02$ & $25.53\pm0.12$ & $0.09\pm0.02$ & $26.48\pm0.23$\\
\hline 
\end{tabular}
\end{table*}
We present new \textit{HST} data taken with the Ultraviolet-Visible (UVIS) channel of the Wide Field Camera 3 (WFC3) in four filters (proposal ID 17290, PI Y.~Chen), two ultraviolet (UV) and two optical filters, corresponding to the filters used for the previous data presented in \cite{Sun2022, Sun2023, Chen2023} and \cite{Inkenhaag2023}. The F225W, F336W and F555W data were taken 1900~days after the first detection of the transients, while the F814W observations failed at that epoch and so were repeated at 2043~days. We take the first detection of the transient as T$_{\rm 0} = 58285.44$ \citep{Perley2019}. All observations were completed using gyro guiding, after the reacquisition of the guide star failed. We provide a summary of the observations in Table~\ref{tab:data}.

To obtain photometry of the new data, we follow the procedure described in \cite{Inkenhaag2023}. In short, this entails combining the individual \texttt{\_flc} images using {\sc astrodrizzle} from the python package {\sc drizzlepack} \citep{drizzlepack}\footnote{\url{https://drizzlepac.readthedocs.io/en/latest/astrodrizzle.html}}. The final pixel scale is decreased to \texttt{final\_scale=0.025} for a better sampled point spread function (PSF). We then align the new images to the images in the same filter taken at 714~days from \citep{Inkenhaag2023}, as the 714~days images were aligned between the different filters and aligning to images in the same filter is more accurate. The images used in \cite{Inkenhaag2023} were aligned at the pixel level, to allow for dual image mode aperture photometry. 
We perform aperture photometry on the images using {\sc Source Extractor} \citep{Bertin1996} in dual image mode. The base image used for source detected is the F336W image at 714~days, and the aperture used is 0.08~arcsec, which correspond to a diameter of $\sim2$ times the Full Width at Half Maximum (FWHM). Enclosed energy corrections are done using the table provided in the WFC3 handbook website\footnote{\url{https://www.stsci.edu/hst/instrumentation/wfc3/data-analysis/photometric-calibration/uvis-encircled-energy}}.

As in \cite{Inkenhaag2023}, we also perform PSF photometry using {\sc dolphot} (v2.0; \citealt{Dolphin2000}), which performs PSF photometry on the individual \texttt{\_flc} frames and combines the individual measurements into a final (Vega) magnitude for each source. We again use the F336W epoch one image as our base image, align the individual \texttt{\_flc} images to this image using {\sc tweakreg} from {\sc drizzlepack}, and use the sources detected in this base image as fixed positions in the ``warmstart" option in {\sc dolphot} for our new images. We use the zeropoints from the WFC3 instrument handbook\footnote{\url{https://www.stsci.edu/hst/instrumentation/wfc3/data-analysis/photometric-calibration/uvis-photometric-calibration}} to convert between Vega and AB magnitudes. 

The results of our aperture and PSF photometry are listed in Table~\ref{tab:data} and we plot the lightcurve including the new data in Figure~\ref{fig:lightcurve}, using the previous data from \cite{Inkenhaag2023}. There is significant fading between the last epoch previously reported in \cite{Inkenhaag2023} and the new data at 1900/2043 days only in the F336W filter. In none of the other filters do we detect evidence for fading at more than 3$\sigma$. 

\section{Discussion} \label{sec:discussion}

\begin{figure}
 \centering
 \hspace*{-.3cm}\includegraphics[width=0.5\textwidth]{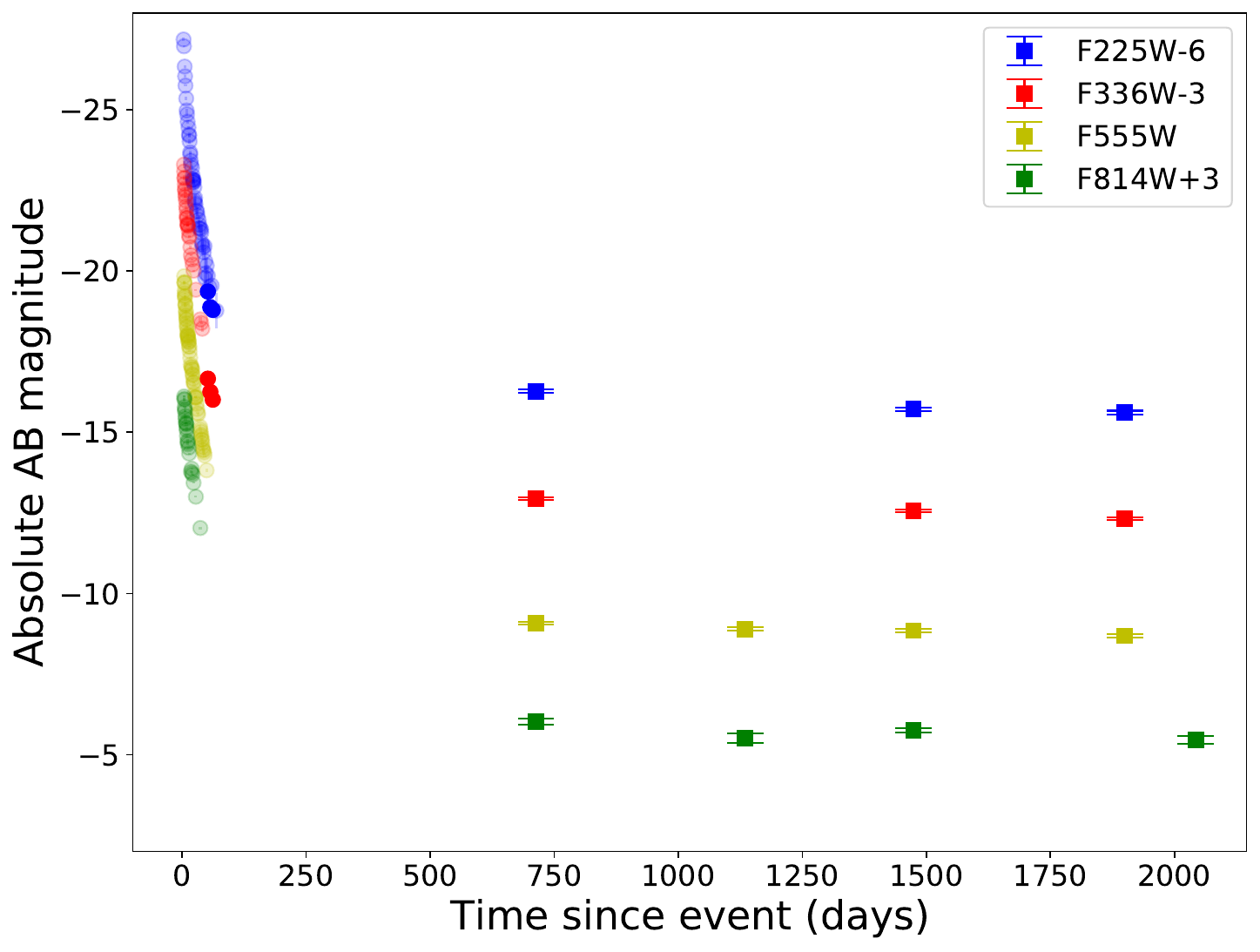}
  \caption{Lightcurves in four \textit{HST} filters from \protect\cite{Inkenhaag2023}, with the data present in this work added (data after t=1750~d is the new data), to show the rate of decay in all four filters slowed down significantly at later times, resembling the lightcurves of TDEs. The lightcurves in the different filters are offset for clarity as indicated in the legend, and different colours correspond to the different filters (F225W in blue, F336W in red, F555W in yellow and F814W in green). We note that the 1$\sigma$ error bars on the magnitudes are small, the horizontal bars through the markers are the endcaps of these error bars. The absolute magnitudes plotted in blue and red (UV filters) are intrinsic to the source, while the magnitudes plotted in yellow and green data (optical filters) likely originate in an underlying extended source \protect\citep{Inkenhaag2023}. }
 \label{fig:lightcurve}
\end{figure}

\begin{figure}
 \centering
 \hspace*{-1.cm}\includegraphics[width=0.55\textwidth]{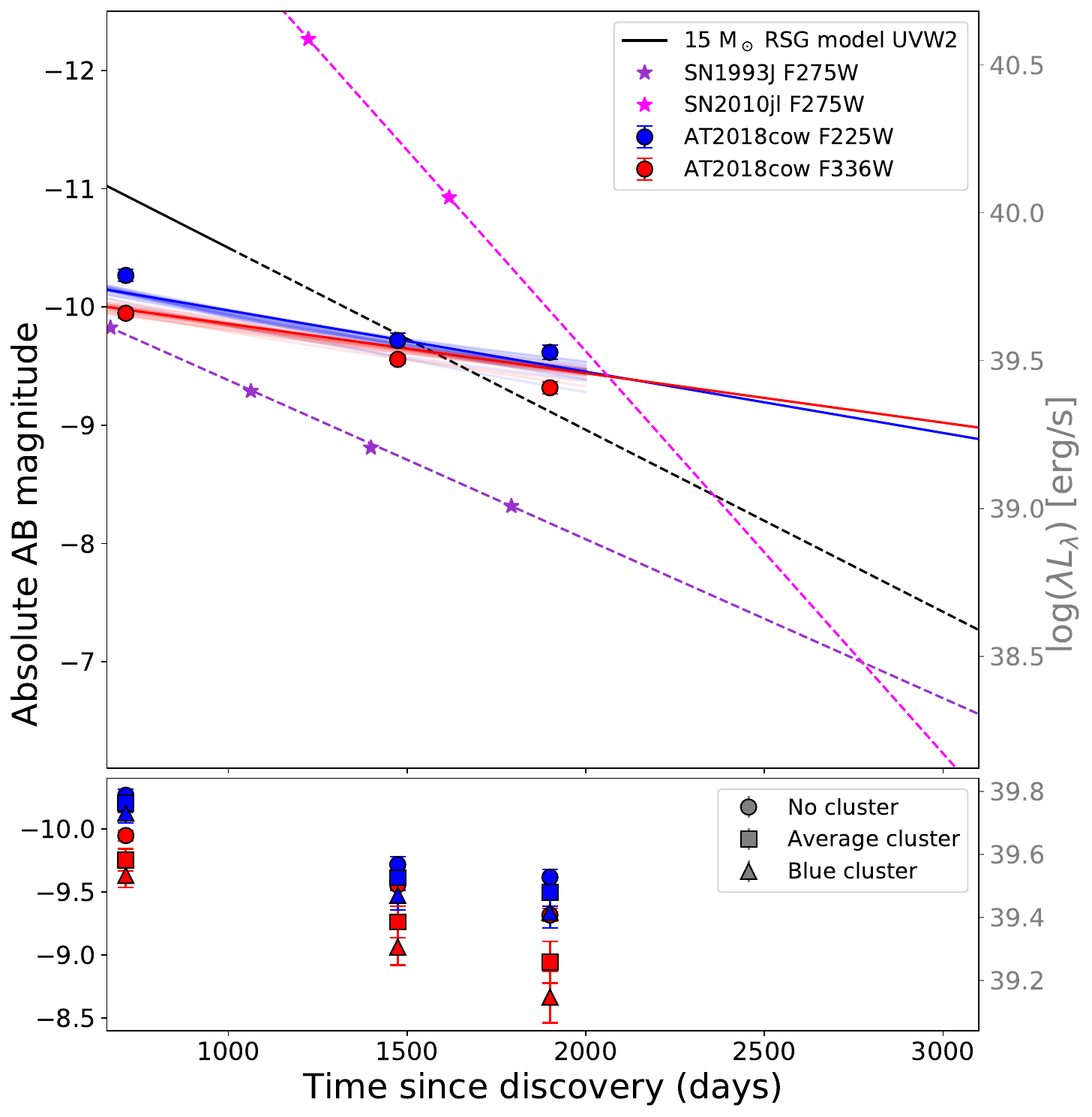}
  \caption{\textit{Top panel:} Late time (>660~days) UV lightcurve of \cow\ in the F225W (blue) and F336W (red) filters, assuming all emission is coming from \cow. The red and blue lines are the disk modelling from \protect\cite{Inkenhaag2023} in these same filters, which were fit to the first two epochs of data shown only and extrapolated out to later times here (no further fitting has been performed). 
  The third epoch at 1900~days is consistent with the predictions from this model, under the assumption that all UV light come from \cow\ (see Section~\ref{sec:discussion}). The purple and pink stars are synthetic photometric measurements in the F275W filter from two interacting supernovae, SN~2010jl and SN~1993J, respectively, as presented in \protect\cite{Inkenhaag2025}. The dashed lines are straight line fits to the magnitudes as a function of time, extrapolated to show the decay rate of in the lightcurve to later times. The black line is the model for an interacting SN from \protect\cite{Dessart2023}, extrapolated after 1000~days. \textit{Bottom panel:} Zoom-in on the lightcurve of \cow, where the circular data points represent the UV magnitudes assuming all the UV light is emitted by the transient, the square markers represent the UV magnitudes when a constant underlying source with the average SEDs of the compact SF regions from \protect\cite{Inkenhaag2023} is subtracted, and the triangular data points represent the UV magnitudes when a constant underlying source with the bluest SED that has F815W-F555W colours redder than \cow\, is subtracted (see the main text for details). Note that the circular markers for the F336W filter are behind the square and triangular markers for the last two epochs. We have also plotted the average disk model for each of the cluster assumptions, which include the data of the third epoch, in the same style as in the top panel.}
 \label{fig:model_comp}
\end{figure}

As was also the case in \cite{Inkenhaag2023}, the {\sc dolphot} PSF photometric measurements are not fully consistent with the aperture photometry, especially in the optical bands, where {\sc dolphot} gives fainter magnitudes than {\sc Source Extractor}. This can be explained by an extended background source at the position of \cow, for which a circular aperture will naturally measure more flux than when fitting with a PSF. The presence of this underlying, extended source was shown to be likely in \cite{Inkenhaag2023}, with its main contribution being in the optical bands. In the UV bands the difference in magnitude between aperture and PSF photometry is smaller. This is because there is still a point source present at the position of \cow, which dominates the flux. Therefore the ratio between the point source flux and the flux of the extended region is much higher, leading to a smaller difference when performing aperture photometry compared to PSF photometry. 

Despite the difference in magnitudes between the methods, when we compare the change in magnitude between epochs 3 and 4, both methods provide similar results. The only band in which there is significant ($>3\sigma$) fading is the F336W band for either method. The other three bands do not display significant fading. For the optical bands, this solidifies the conclusion from \cite{Inkenhaag2023} that we are likely observing a spatially extended, underlying source in those bands. The significant change in F336W flux also supports the hypothesis that we are still observing the slowly fading transient in the UV bands. It does raise the question why there is only significant fading in the F336W filter and not in the F225W filter, as for any cooling, which is perhaps expected for an accretion disk, we would also expect decay in the F225W filter if decay in the F336W filter is observed.
This could be because the fading is too slow to detect in the $\sim400$~days since the last epoch, and small amounts of stochastic noise in either observation (coupled to a very slow intrinsic decay) could prevent the detection of significant changes in flux. 

As can be seen from the lightcurves in Figure~\ref{fig:lightcurve}, the \cow\ UV plateau in the lightcurve of two UV bands is even clearer now than it was at 1474~days. This plateau behaviour is also seen in all ($N=40$) TDEs at $z<0.07$ with more than three years of data available \citep{Mummery_et_al_2024, MummeryVV24}, including those TDEs around lower mass SMBHs \citep[<$10^{6.5}$M$_\odot$;][]{VanVelzen2019, MummeryVV24}. 

Additionally, and crucially, one can see from Figure~\ref{fig:model_comp} that the newest data points are entirely consistent with the predictions from the disk modelling presented in \cite{Inkenhaag2023}. This model is premised  upon an accretion disk that forms close to the black hole which then cools down and spreads out (to conserve angular momentum) when mass is accreted onto the compact object. This spreading out creates a larger surface area of the accretion disk, which compensates for the cooling, creating a (near-) constant luminosity in the UV. This constant luminosity is what we observed as the UV plateau in the lightcurve. There is no external source of material being added onto the accretion disk in this model and therefore it is not analogous to either an AGN disk or a high-mass X-ray binary disk, it has to be an isolated object spontaneously forming the accretion disk close to the black hole, which we believe is most likely to be a TDE\footnote{The {\tt FitTeD} code behind the disk model used to make these plots is publicly available \citealt{mummery2024fitted}.}. We stress that no fitting has been performed to these final data points, and the curves seen in Figure~\ref{fig:model_comp} simply represent the extrapolation of the \cite{Inkenhaag2023} models to later times. Combined, we argue that this strongly supports the model that \cow\ originates from a TDE. 

As a sanity check, we have also reperformed the disk modeling including the third epoch of data. We find a mass of $\log(M_{\rm BH})=3.3\pm0.4$, which is fully consistent with the black hole mass presented in \cite{Inkenhaag2023}, with a smaller uncertainty. While we believe the consistency of the new observations with the predictions of the previously presented model is perhaps a more novel argument, the mass of the disk model including the new data points is also support for \cow\ originating from a TDE.

\begin{table}
\caption{Magnitudes (AB) at the position of \cow\ after subtraction of various assumptions for the SED of an underlying stellar cluster. See the main text for how we obtain these average SEDs and resulting magnitudes. The resulting mass of the black hole assuming a TDE origin is also listed, the lightcurves for the various disk model fits can be found in Figure~\ref{fig:SEDsub_models}.
}
\label{tab:cluster_sub}
\begin{tabular}{lccc}
\hline
Epoch & F225W & F336W & log($M_{BH}$/M$_\odot$) \\
(days) & (mag) & (mag) & \\
\hline
\textbf{No cluster} & & & 3.3$\pm$0.4\\
\hspace{3mm} 714 & 23.73$\pm$0.05 & 24.05$\pm$0.04 & \\
\hspace{3mm} 1274 & 24.28$\pm$0.06 & 24.44$\pm$0.04 & \\
\hspace{3mm} 1900 & 24.38$\pm$0.06 & 24.68$\pm$0.05 & \\
\textbf{Average cluster}  & & & 3.0$\pm$0.5 \\
\hspace{3mm} 714 & 23.79$\pm$0.07 & 24.24$\pm$0.08 & \\
\hspace{3mm} 1274 & 24.38$\pm$0.10 & 24.73$\pm$0.12 & \\
\hspace{3mm} 1900 & 24.50$\pm$0.11 & 25.05$\pm$0.16 & \\
\textbf{Blue cluster} & & & 3.1$\pm$0.5 \\
\hspace{3mm} 714 & 23.87$\pm$0.08 & 24.37$\pm$0.10 & \\
\hspace{3mm} 1274 & 24.52$\pm$0.11 & 24.93$\pm$0.14 & \\
\hspace{3mm} 1900 & 24.66$\pm$0.12 & 25.33$\pm$0.20 & \\
\hline 
\end{tabular}
\end{table}

We investigate the possible influence of an underlying cluster contributing to the emission in the UV filters by subtracting the predicted UV flux for an underlying source using the spectral energy distributions (SEDs) from \citep{Inkenhaag2023} at 713~days. First, we average all SEDs of the compact star forming regions plotted in their figure~ 3, and normalise this to the flux in the F814W filter at our latest epoch. We then subtract the predicted flux in the UV and calculate the corresponding magnitude of the transient. We repeat the process with the SED that has the bluest F814W-F225W colour, but also has a redder F814W-F555W colour than \cow\ to make sure the predicted F555W flux does not exceed the detected F555W flux. We plot these magnitudes in Figure~\ref{fig:model_comp} as well, with square and triangular markers for the average SED subtraction and the bluest SED subtraction, respectively, and list the magnitudes in Table~\ref{tab:cluster_sub}. As expected, the decay of \cow\, in the UV becomes faster the bluer the underlying cluster we assume, as it would be contributing more constant flux in the UV bands. Therefore, a steeper decay of the transient would be needed to accommodate the small fractional decay in UV flux observed. We do note that the uncertainties on the magnitudes after subtracting the predicted UV flux are significantly larger than when no underlying cluster is assumed due to the uncertainty in the normalisations of the assumed clusters, to the point that the magnitudes are all still consistent within 3$\sigma$. Nevertheless, the true rate of fading of the transient is expected to lie between the two extremes (no UV contribution of an underlying star cluster -- the bluest star cluster that is consistent with all the UV-optical data). 

\begin{figure}
 \centering
 \hspace*{-.55cm}\includegraphics[width=0.55\textwidth]{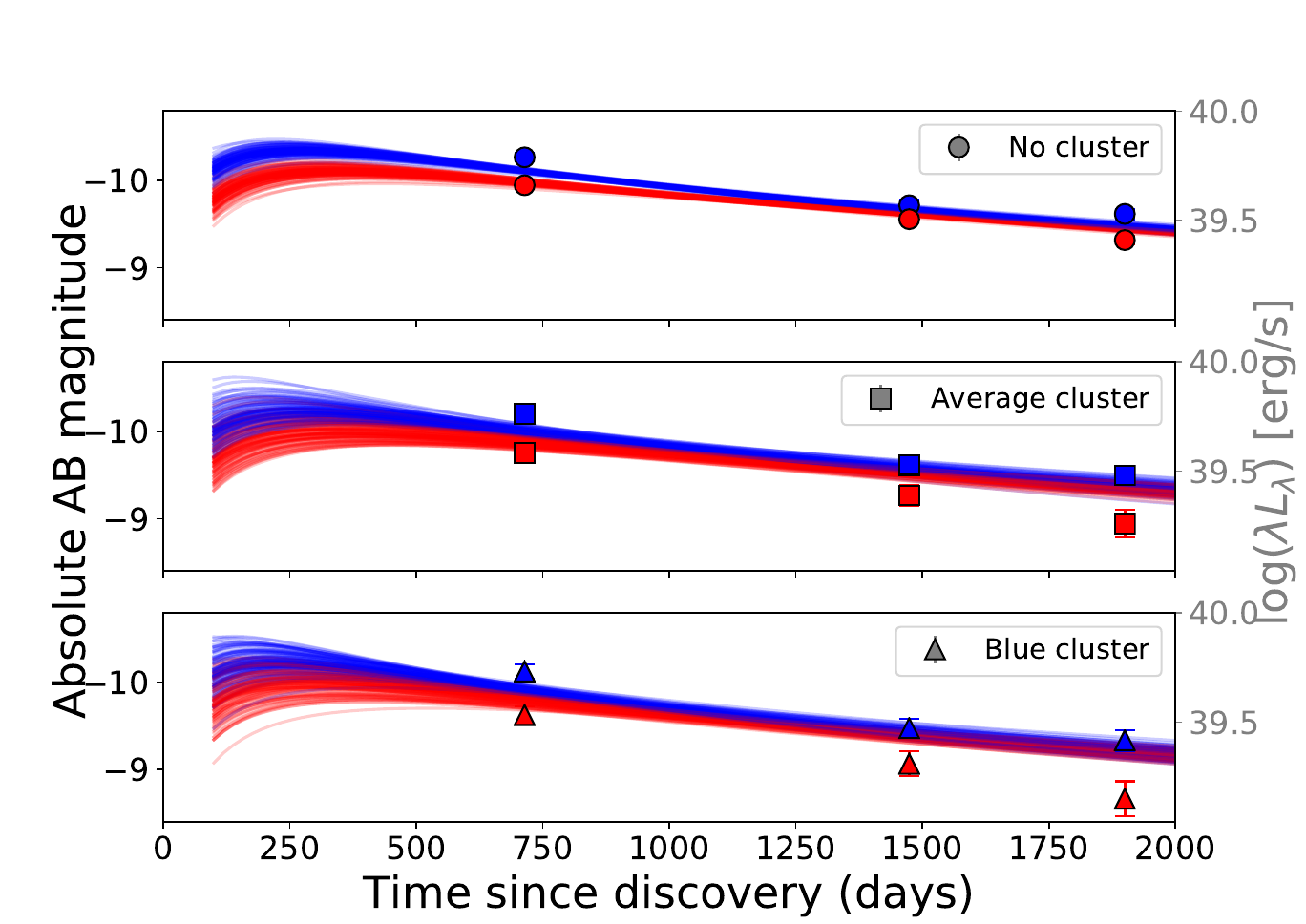}
  \caption{Late time (>660 days) UV lightcurves of \cow\ assuming no underlying cluster (\textit{top panel}), an average cluster obtained from the compact SF regions in the host galaxy (\textit{middle panel}) or a blue cluster (\textit{bottom pabel}), see the main text for details. F225W magnitudes are plotted in blue and F226W magnitudes are plotted in red. The lines in each panel are 100 disk models for each filter, from which the mass of the BH has been deduced, with the colours corresponding to the filter. These disk models are the models from \protect\cite{Inkenhaag2023}, but repeated to include the third epoch in the fit. We find BH masses in each case which are consistent with the BH mass found in \protect\cite{Inkenhaag2023}. 
  }
 \label{fig:SEDsub_models}
\end{figure}

To explore how the underlying cluster might influence the parameters of the disk model, we also run the disk model for the scenarios were we assume an average and a blue underlying stellar cluster and plot these in Figure~\ref{fig:SEDsub_models}. As can be seen from Table~\ref{tab:cluster_sub}, the black hole mass shifts by $\sim0.3$~dex between the no underlying cluster and the average/blue underlying clusters scenarios. However, the uncertainty on the black hole mass is of the order of $\sim0.5$~dex, which means both scenarios are still consistent with the decay rate for a TDE disk model with a black hole mass which is consistent with that presented in \cite{Inkenhaag2023}.

While the disk model presented in \cite{Inkenhaag2023} (and reproduced here) assumed that the disk formed from the tidal disruption of a low-mass star, we stress that the late-time UV plateau luminosity is only a weak function of stellar properties \citep{Mummery_et_al_2024}, meaning that a white dwarf tidal disruption (which may well provide a natural explanation for the rapid early light curve decay) is also consistent with this late time UV emission, and would require only a slightly higher black hole mass than presented in \cite{Inkenhaag2023}. If a much higher mass star was involved in the disruption \citep[such as a Wolf-Rayet star as has been suggested by][]{Metzger22b} then a less massive black hole would be required. High disk masses ($M_{\rm disk}\sim 20M_\odot$) would be required for the central black hole in \cow\ to be at the scale of a typical X-ray binary black hole $(M_{\rm BH} \sim 20 M_\odot$), although modelling a disk system with mass comparable to the central object is highly non-trivial, and beyond the scope of this paper.

\cite{Inkenhaag2025} investigated the possibility that the late time emission of \cow\ is from an interaction between SN ejecta and a (dense) CSM. There is a lack of modelling of SN lightcurves and spectra beyond $\sim300$~days in the UV for interacting SNe, and the available data for interacting SNe at the timescales of the latest observations of \cow\ are very limited \citep{Inkenhaag2025}. \cite{Khatami2024} model various configurations of CSM out to timescales similar to the observations of \cow, though clearly a special ejecta and CSM geometry is needed to produce a lightcurve that has a plateau phase at later times. In Figure~\ref{fig:model_comp} we have linearly extrapolated the synthetic photometry of two known interacting SNe out to later times. Comparing this to the model from \cite{Dessart2023} of a 15.2~M$_\odot$ red super giant  exploding with injected energy, representing the interaction with the CSM, we can see that observations and modelling yield a range of decay rates for CSM interaction. Despite variation in these decay rates, the lightcurve decay in interacting SNe is generally much faster (and they typically are much fainter at $\Delta t \sim 2000$ days) than the predictions for the TDE disk model, even when subtracting a blue underlying source, adding to the evidence for a TDE nature of \cow. 

We can also see from Figure~\ref{fig:model_comp} the variety in magnitudes from observations and modelling for CSM interaction. The late time UV observations of \cow\ are consistent in magnitude with the upper range of these magnitudes. At even later times, out to $\sim3000$~days, we see a significant difference in magnitude between emission from even the brightest SN--CSM interactions and an accretion disk from a TDE. We will therefore need (at least) another observation on this time scale to fully distinguish between the models.

\section{Summary and Conclusion} \label{sec:conclusions}

We present a new epoch of late-time \textit{HST} observations of \cow. There is significant fading since the last epoch only in the F336W filter, while fading in the F225W, F555W and F814W filters is marginal. The fading in UV between the previous epoch and the epoch presented in this paper is consistent with the fading predicted by the TDE disk model presented in \cite{Inkenhaag2023} (and the BH mass is consistent with previous modelling when including this latest datapoint) and is significantly slower than observed in interacting SNe. The limited available simulations of the UV brightness of interacting CCSNe, which we need to extrapolate to the times observed in this work, also fade faster than what we observe for \cow. However, purely based on brightness, we cannot rule out CSM interaction just yet, especially if we assume a blue underlying star cluster.

Nevertheless, we conclude that the late time UV emission of \cow\ is an argument for a TDE nature of this event, supported by the latest epoch being consistent with model predictions. A final data point when TDE and CSM-interaction models and observations have diverged significantly, around 2800--3000~days, is needed to fully rule out (on brightness arguments) SN--CSM interaction as the nature of the late-time emission.

\section*{Acknowledgements}

The scientific results reported on in this article are based on data obtained under \textit{HST} Proposals 17290 with PI Y.~Chen.
AI acknowledges support from the UK Science and Technology Facilities Council, grant reference ST/X001067/1. This work was supported by a Leverhulme Trust International Professorship grant [number LIP-202-014]. 
This work makes use of Python packages {\sc numpy} \citep[v1.21.6;][]{2020Natur.585..357H}, {\sc matplotlib} \citep[v3.3.4;][]{2007CSE.....9...90H} and {\sc drizzlepac} \citep[v3.1.6;][]{drizzlepack}.
This work made use of Astropy (v4.3.1; \url{http://www.astropy.org}): a community-developed core Python package and an ecosystem of tools and resources for astronomy \citep{astropy:2013, astropy:2018, astropy:2022}.

\section*{Data Availability}

All data used in this paper is publicly available from the \textit{HST} data archive.



\bibliographystyle{mnras}
\bibliography{references} 








\bsp	
\label{lastpage}
\end{document}